\documentclass[twocolumn,secnumarabic,amssymb, nobibnotes, aps,
prd]{revtex4} \usepackage{graphicx}

\begin{document}
\title{Quantum limits to the second law and breach of symmetry}
\author{Alexey Nikulov}
%\email[]{nikulov@ipmt-hpm.ac.ru}
\affiliation{Institute of Microelectronics Technology and High Purity Materials, Russian Academy of Sciences, 142432 Chernogolovka, Moscow District, RUSSIA. nikulov@ipmt-hpm.ac.ru}
%\date{}
\begin{abstract}Connection between an intrinsic breach of symmetry of equilibrium motion and violation of the second law is accentuated. An intrinsic breach only of clockwise - counter-clockwise symmetry of a circular equilibrium motion can be logical under equilibrium conditions, whereas a breach of right-left symmetry should be always an actual violation of the second law. The reader's attention is drawn to experimental evidence of an intrinsic breach of the clockwise - counter-clockwise symmetry of a circular equilibrium motion, well known as the persistent current. The persistent current is observed in mesoscopic normal metal, semiconductor and superconductor loops and the clockwise - counter-clockwise symmetry is broken because of the discrete spectrum of the permitted states of quantum charged particles in a closed loop. The quantum oscillations of the dc voltage observed on a segment of an asymmetric superconducting loop is experimental evidence of the intrinsic breach of the right-left symmetry and an actual violation of the second law.  


\end{abstract}

\maketitle

\narrowtext

%\footnote{ The Invited Lecture of the Conference "Frontiers of Quantum and Mesoscopic Thermodynamics" 26-29 July %2004, Prague,  http://www.fzu.cz/activities/conferences/fqmt04/} 

\noindent
\textbf{1. Introduction}

\bigskip

Arthur Eddington wrote \cite{1}: ``\textit{The second law of thermodynamics holds, I think, the supreme position among the laws of Nature. If someone points out to you that your pet theory of the universe is in disagreement with Maxwell's equations - then so much the worse for Maxwell's equations. If it is found to be contradicted by observation, well, these experimentalists do bungle things sometimes. But if your theory is found to be against the second law of thermodynamics I can give you no hope; there is nothing for it but collapse in deepest humiliation.}'' Therefore, most scientists distrust \cite{2} the challenges to the absolute status of the second law published in recent years [3-26]. The second law was, and is, a subject of belief rather than of knowledge. Almost all people are fully confident that the second law cannot be violated, although nobody can explain why. 

The belief in the second law is based upon, and is inseparably linked with, the centuries-old belief in impossibility of a perpetuum mobile. Carnot's principle, which we call since Clausius time \textit{ "the second law of thermodynamics''} \cite{27}, was postulated in 1824 on the basis of this belief. It was obvious already in Carnot's time that in order for a perpetuum mobile to be impossible, an irreversible behavior should exist in nature. A perpetuum mobile would be inevitable, however, according to the law of energy conservation, if all physical processes could be reversible. Therefore, irreversibility is the main feature of thermodynamics, and the second law is first among all statements on irreversibility. 

The demand of irreversibility caused the well known collision between dynamics and thermodynamics. This collision is not overcome completely even to the present time. But, it is interesting that most scientists rejected in the 19th century the atomic-kinetic theory of the heat proposed by Maxwell and Boltzmann \cite{27} because of this collision, whereas in the 20th century most scientists believed that this theory had eliminated this collision. This belief has remained invariable since the time of Maxwell. The words by J.C. Maxwell written in 1878 [28]: \textit{"the second law is drawn from our experience of bodies consisting of an immense number of molecules''} remains the only scientific substantiation of the second law up to the present time [29]. 

This probabilistic substantiation looks very convincing: an uncompensated decrease of entropy is improbable since a macroscopic system moves spontaneously from a less probable state to a more probable state in the majority of cases, and this majority becomes so overwhelming when the number of atoms in the system becomes very large that irreversible behavior becomes a near certainty [29]. But this obvious mathematical argument is not enough, because of the existence of perpetual motion, i.e. the motion under equilibrium conditions of atoms, molecules, small Brownian particles and others. It is, therefore, necessary to postulate absolute randomness of any equilibrium motion in order to save the second law. For example, the second law could be broken even without a ratchet and pawl in the system, as was considered by Feynman [30] (and earlier by Smoluchowski \cite{27}) if the average velocity of molecules is not zero under equilibrium conditions. 

The randomness of any equilibrium motion was postulated by Maxwell and Boltzmann, and this postulate is used through the present time as being self-evident, without any substantiation. Physics is an empirical science. But, physical knowledge is based not only on empirical data but also on postulates which seem self-evident. It is important to emphasize a connection of the randomness postulate with symmetry, which seems self-evident, also. An existence of any equilibrium motion with non-zero average velocity means an intrinsic breach of symmetry. For example, circular motion breaks clockwise - counter-clockwise symmetry and a directed motion breaks right - left symmetry. An intrinsic breach of symmetry is a very grave matter. Nevertheless, it is observed in full accordance with the bases of quantum physics. The purpose of the present paper is to draw reader's attention to this obvious fact, and to violation of the second law that is connected with it. 

\bigskip

\noindent
\textbf{2. Symmetry saves the second law }

\bigskip

There is a logical contradiction to the possibility of an actual violation of the second law under equilibrium conditions. An actual violation of the second law means systematic reduction of the total entropy $dS/dt < 0$, whereas the total entropy cannot change under equilibrium conditions $dS/dt = 0$. Therefore, an actual violation of the second law means violation of the equilibrium state. 

There is a connection of this contradiction with a logical possibility of an intrinsic breach of different types of symmetry. The energy should move from one part to another part of the system, i.e., the right - left symmetry should be broken with an actual violation of the second law, whereas only clockwise - counter-clockwise symmetry of circular motion can be broken under equilibrium conditions, since the breach of right - left symmetry of any motion means an irreversible transference of mass or energy in violation of equilibrium. 

It is well known that any element of an electric circuit can be a power source at a finite temperature $T$ because of equilibrium noise \cite{30}, as has been described theoretically by Nyquist [31] and observed by Johnson [32] as long ago as 1928. But this power cannot be used under equilibrium conditions since the power of each element is distributed among the same frequency $\omega $ spectrum: $W_{Nyq} = k_{B}T\Delta \omega $ from $\omega  = 0$ up to the quantum limit $\omega  < k_{B}T/\hbar  $. There is symmetry under equilibrium conditions. We can not say what element is the power source and which element is the load. This right - left symmetry is broken when equilibrium conditions are broken, i.e., the elements are under different temperatures $T_{1} > T_{2}$. Then the element at $T_{1}$ is the power source and the other one at $T_{2}$ is the load. The power source and the load can be distinguished at $T_{1} = T_{2 }$ only if their frequency spectrums are different. 

The equality of the power $W_{Nyq} = k_{B}T\Delta \omega $ at any frequency is a consequence of the randomness of the equilibrium motion and intrinsic symmetry. Any distinction of the frequency of equilibrium motion from other frequencies can only be at the violation of this symmetry. Thus, we may say that the intrinsic symmetry saves the second law against our experiencing perpetual motion, and violation of the second law can be possible only if an intrinsic breach of symmetry is observed. 

\bigskip

\noindent
\textbf{3. Difference between external and intrinsic breach of symmetry }

\bigskip

Most of the numerous challenges to the second law are connected with attempts to break symmetry and to order random equilibrium motion. The ratchet/pawl combination \cite{30} is best known, and the rectification of the Nyquist noise [33] with help of diodes is most popular. Feynman \cite{30} and others [34] have shown that the second law wins in this fight. The ratchet/pawl combination and the diode break symmetry, but this breach is external, i.e., it is not based on a natural law but is man-made, and therefore cannot put order into random equilibrium motion. The second law wins since the pawl undergoes random Brownian motion and the frequency spectrum of the equilibrium power of the diode does not differ from the other element of electric circuit. 

The postulate is that randomness of any equilibrium motion can be violated only by an intrinsic breach of symmetry, i.e., a breach based on a natural law, which governs, but does not undergo, the equilibrium motion. All material objects undergo thermal equilibrium motion. Therefore, the fight against the second law with help of a man-made breach of symmetry is a failure. 

\bigskip

\noindent
\textbf{4. Validity of the randomness postulate according to classical mechanics}

\bigskip

Equilibrium motion is not ordered, i.e., the average velocity of equilibrium motion equals zero $<v> = 0$ and the equilibrium power is the same $W_{Nyq} = k_{B}T\Delta \omega $ at any frequency (in the classical limit), because of symmetry and homogeneity of space and homogeneity of time. Nobody can doubt these postulates of physical knowledge. Because of space symmetry, the probability $P$ of a microscopic state of a particle does not depend on the velocity direction $P(v) = P(-v) = P(v^{2})$. Therefore, $<v> = \sum_{per.st.} v P(v^{2}) + (-v) P(v^{2}) = 0$ when a permitted state with opposite velocity $-v$ exists for any permitted state with a velocity $v$. All states are permitted according to classical mechanics. Consequently, Maxwell and Boltzmann could substantiate the randomness postulate which they used. 

\bigskip

\noindent
\textbf{5. Intrinsic breach of symmetry because of discreteness of permitted state spectrum }

\bigskip

But according to quantum mechanics, a spectrum of permitted states can be discrete. It is important that in the discrete spectrum of the momentum circulation 
$$\oint_l {dlp} = \oint_l dl(mv + qA) = m\oint_l dlv + q\Phi = n2\pi \hbar   \eqno{(1)}$$
the state with zero velocity $v = 0$ is forbidden when the magnetic flux $\Phi $ inside the path $l$ of circulation (1) is not divisible by the value $\Phi_{0} = 2\pi \hbar  /q$ called the flux quantum $\Phi  \neq n\Phi _{0}$ and that the state with opposite velocity $v$ and $-v$ cannot be permitted simultaneously at $\Phi \neq n\Phi _{0}$ and $\Phi \neq (n+0.5)\Phi _{0}$. The average value of the velocity circulation 
$$\oint_l {dlv} = \frac{2\pi \hbar }{m}(n - \frac{\Phi }{\Phi _0 }) \eqno{(2)}$$
of a quantum particle with a charge $q$ can be non-zero under equilibrium conditions at $\Phi \neq n\Phi _{0}$ and $\Phi \neq (n+0.5)\Phi _{0}$ because of the features of quantum physics (1) and of the momentum of the charged particle $p = mv + qA$ including not only velocity but also the vector potential $A$. 

The permitted state with lowest energy $E$ has the highest probability $P \propto exp(-E/k_{B}T)$. The permitted state (2) with lowest kinetic energy $E_{kin} = mv^{2}/2$ can have an opposite direction of the velocity $v$ at different $\Phi /\Phi _{0 }$ values: for example if this direction is clockwise at $\Phi /\Phi_{0} = 1/4$ when $min(v^{2}) \propto min(n - \Phi /\Phi _{0})^{2} = (-1/4)^{2}$ at $n = 0$ then it is counter-clockwise at $\Phi /\Phi_{0 } = 3/4$ when $min(v^{2}) \propto min(n- \Phi /\Phi _{0})^{2} = (1/4)^{2}$ at $n = 1$. Therefore, the average velocity $<v> = \sum_{per.st.} v P(v^{2})$ in the equilibrium state should be a periodical function of the magnetic flux and its direction changes with the scalar value $\Phi /\Phi_{0}$ without any vector factor when the energy difference between adjacent permitted states (2) is higher than the energy of thermal fluctuations $E_{kin}(n+1) - E_{kin}(n) = mv^{2}(n+1)/2 - mv^{2}(n)/2 \approx \hbar ^{2}/2mr^{2} > k_{B}T$. The latter can only occur at a very low temperature in the case of the quantization of single electron states in a structure with a radius $r$ accessible for current nano-technology, for example, $\hbar ^{2}/2mr^{2}  \approx k_{B} \ 0.01 \ K$ at $r = 0.5 \ \mu m$.

\bigskip

\noindent
\textbf{6. Experimental evidence of the intrinsic breach of clockwise - counter-clockwise symmetry}

\bigskip

Nevertheless, the persistent current $j_{p} = qn_{q}<v>$ with a periodical dependence $j_{p}(\Phi /\Phi_{0})$ has been observed even in a normal metal [35] and in semiconductor mesoscopic structures [36], first observed in 1990. But the first experimental evidence of the intrinsic breach of clockwise - counter-clockwise symmetry was obtained on a superconducting structure [37].

Superconductivity is a macroscopic quantum phenomenon, since superconducting pairs have the same value $n$ of the momentum circulation (1) and the energy difference between adjacent permitted states (1) should be multiplied by the number $N_{s} = V_{s}n_{s}$ of pairs in the superconductor: $E_{kin}(n+1) - E_{kin}(n) \approx  N_{s}\hbar ^{2}/2mr^{2} \gg  k_{B}T$ in any real case. Therefore, the first experimental evidence of the persistent current in a superconductor, the Meissner effect, was discovered as long ago as 1933. The Meissner effect is observed in a bulk superconductor, wherein in the interior thereof $\Phi  = 0$ since the velocity $v_{s} = 0$ along a closed path $l$ and $n = 0$, see (1), the wave function of the superconducting pairs does not have singularity inside $l$. The second and third experimental evidence of the persistent current, flux quantization and velocity quantization (2) were obtained in 1961 [38] and 1962 [39]. The quantization of the magnetic flux is observed at a strong screening in thick-walled superconducting cylinder or loop where $v_{s} = 0$ along a closed path $l$ and therefore $\Phi  = n\Phi_{0}$ according to (2). 

The periodic variation of the velocity (2) $<v_{s}(\Phi /\Phi_{0})>$ is observed in thin-walled cylinder [39] or loop [40] with weak screening $LI_{p} < \Phi_{0}$. The persistent current $I_{p}(\Phi/\Phi_{0}) = sj_{p} =s2en_{s} <v_{s}>  \propto  <n> - \Phi/\Phi_{0}$ in loops both with and without Josephson junctions, where the thermodynamic average value $<n>$ of the quantum number $n$ is close to an integer number $n$ corresponding to minimum energy, i.e., to the minimum of $(n - \Phi /\Phi_{0})^{2}$, when the magnetic flux $\Phi  = BS + LI_{p} \approx BS$ inside the loop is not close to $(n + 0.5)\Phi _{0}$. The persistent current $I_{p}(\Phi /\Phi_{0})$ can be observed in a superconducting loop even with very long length of the circumference $l = 2\pi r$ and very small cross-section $s$ since the density of the superconducting pairs, $n_{s}  \approx 10^{28} \ m^{ - 3}$ for $T \ll  T_{c}$, is very great: for example $N_{s}\hbar ^{2}/2mr^{2} = \pi sn_{s}\hbar^{2}/mr  \approx k_{B} 60 \ K$ at $l = 2\pi r = 10 \ m$ and $s =1 \ \mu m^{2}$.

The intrinsic breach of clockwise - counter-clockwise symmetry is observed in the closed superconducting state, when the density of the superconducting pairs $n_{s} > 0$ along the whole of the loop circumference $l$, $I_{p} \neq 0$, $R = 0$. The  $n_{s }$ should be replaced in the relation for energy difference $E_{kin}(n+1) - E_{kin}(n)$ between adjacent permitted states (2) by the average value 
$$ < n_s^{ - 1} > ^{ - 1} = (l^{ - 1}\oint_l {dl} n_s^{-1} )^{-1}\eqno{(3)}$$
when the density of the pairs is not homogeneous along the loop circumference $l$ since the persistent current $I_{p} =sj_{p} = s2en_{s}v_{s}$ should be constant along $l$ in the stationary case. The spectrum is strongly discrete $E_{kin}(n+1) - E_{kin}(n) = \pi s<n_{s}^{-1}>^{-1}\hbar ^{2}/mr \approx k_{B} 60 \ K$, the equilibrium velocity $<v_{s}> =\hbar /mr)(<n> - \Phi /\Phi_{0}) \approx \hbar /mr)(-1/4) \approx 10^{-5} \ m/s$ and the persistent current with a density $j_{p} = 2en_{s}<v_{s}>  \approx 3 \ 10^{4} \ A/m^{2}$ can be observed in a loop $l = 2\pi r = 10 \ m$ with a clockwise direction (for example) at $\Phi/\Phi_{0} = 1/4$ and counter-clockwise direction at $\Phi /\Phi_{0 }=3/4$ when $<n_{s}^{-1}>^{ - 1} = n_{s} \approx 10^{28} \ m^{ - 3}$. But the spectrum becomes continuous and the intrinsic breach of symmetry is absent for even a very short segment $l_{s}$, for example with $l_{s} \approx 1 \ \mu m = l \ 10^{-7}$, is switched in the normal state since $<n_{s}^{ - 1}>^{ - 1 } = 0$, $I_{p} = 0$, $R > 0$ when $n_{s} = 0$ in any loop segment. 

Superconducting pairs in a whole long loop are braked at $n_{s} = 0$ in $l_{s}$ down to zero during the time of current relaxation $\tau_{RL} = L_{l}/R$ because of pure classical electric force $mdv_{s}/dt = 2eE = 2e\nabla V$. Where $V(t) = R_{ls}I(t) = R_{ls}I_{p} \exp(-t/\tau_{RL})$ is the potential difference because of a non-zero resistance $R_{ls} > 0$ of the $l_{s}$ segment in the normal state; $L_{l}$ is the inductance of the loop $l$. The opposite change from $<v_{s}> = 0$ to $<v_{s}> = (\hbar /mr)(<n> - \Phi /\Phi_{0}) \neq 0$ takes place because of the quantization (1), without any classical force \cite{41} when the $l_{s}$ segment returns to the superconducting state. It is a manifestation of the well known difference between a superconductor and a classical conductor with infinite conductivity. According to classical mechanics, the momentum circulation should remain invariable without any force, whereas according to quantum mechanics the momentum circulation should be equal to the quantum value $n2\pi \hbar $ where the integer number $n$ can change without any classical force. Therefore, superconducting pairs accelerate against the Faraday electric force when a superconductor expels magnetic field at the Meissner effect. 

The momentum circulation (1) of pair changes between quantum value $n2\pi \hbar $ and $q\Phi  = 2e\Phi $ corresponding $v_{s} = 0$ when the loop is switching between the superconducting states with different connectivity of the wave function, 
i.e., between $n_{s} = 0$ and $n_{s} > 0$ in $l_{s}$. The change of the momentum circulation because of the quantization in time unity equals $(<n>2\pi \hbar  - 2e\Phi )\omega_{sw} = 2\pi \hbar (<n>- \Phi /\Phi_{0})\omega_{sw}$ at the switching frequency $\omega_{sw } \ll  1/\tau_{RL}$. The value $F_{q} =2\pi \hbar (<n> - \Phi /\Phi_{0})\omega_{sw}/l$ was called in \cite{23} \textit{quantum force}. Clockwise or counter-clockwise direction of the quantum force, as well as of the persistent current, is determined with the scalar value $\Phi /\Phi_{0}$ without any vector factor.

\bigskip

\noindent
\textbf{7. Challenge to the second law}

\bigskip

The persistent current is observed \cite{39,40} and predicted \cite{42} even in the fluctuation region near $T \approx  T_{c}$ and above $T > T_{c}$ superconducting transition where the resistance of superconducting loops is not zero, $R_{n} >  R > 0$. The first experimental evidence of $I_{p} \neq 0$ at $R > 0$ is the Little-Parks oscillations of the resistance $R(\Phi /\Phi_{0})$ observed as early as 1962 \cite{39}. The Little-Parks oscillations \cite{39,40} and the observations of $I_{p}(\Phi /\Phi_{0})$ in normal metal and semiconductor mesoscopic loops \cite{35,36} are experimental evidence of a persistent current, i.e., the equilibrium undamped direct current $I_{p} \neq 0$, observed at non-zero dissipation: $RI_{p}^{2} > 0$ at $R > 0$ and $I_{p} \neq  0$. Any undamped current can be observed at non-zero power dissipation only if a power source maintains it. Therefore, the observation \cite{35,36,39,40} of $I_{p} \neq 0$ at $R > 0$ is experimental evidence of a source of persistent power, i.e. of a dc power $W_{p} =RI_{p}^{2}$ existing under equilibrium conditions. 

The persistent power $W_{p} =RI_{p}^{2}$ is a fluctuations phenomenon, like the Nyquist's noise. It is most obvious in the case of a superconducting loop \cite{40} where $RI_{p}^{2} \neq  0$ is observed only in the fluctuation region near $T_{c}$. Above this region $R > 0$ but $I_{p} = 0$ and below it $I_{p} \neq 0$ but $R = 0$. The persistent current $I_{p}  \neq 0$ and the resistance $R > 0$ are non-zero near $T_{c}$ since thermal fluctuations switch the loop between superconducting states with different connectivity \cite{23}. The persistent current does not go out slowly since the dissipation force is compensated by the quantum force, i.e., by the momentum change because of the quantization \cite{23}. 

The persistent current at $R >0$ is Brownian motion like the Nyquist's noise current $<I_{Nyq}^{2}> = k_{B}T\Delta \omega /R$ in a loop with a continuous spectrum. The maximum power of the persistent current $W_{p} = RI_{p}^{2} < (k_{B}T)^{2}/\hbar $ \cite{23} and the total power of the Nyquist's noise are close to the power of thermal fluctuations $W_{fl} = (k_{B}T)^{2}/\hbar $. But there is an important difference between these two fluctuation phenomena. The power of the Nyquist's noise is "spread" $W_{Nyq} = k_{B}T\Delta \omega $ in the frequency region from zero $\omega = 0$ to the quantum limit $\omega  = k_{B}T/\hbar $, whereas the power of the persistent current is not zero at the zero frequency band $\omega  = 0$. 

The difference of the frequency spectrum of the equilibrium persistent power from the equilibrium power of the Nyquist's noise is the consequence of the intrinsic breach of the clockwise - counter-clockwise symmetry and it can break the symmetry of the elements of an electric circuit under equilibrium conditions. Therefore, the persistent power provides a potential possibility for a violation of the second law. The Nyquist's noise is chaotic Brownian motion \cite{30} and the persistent current at $R > 0$ is ordered Brownian motion \cite{23}. Therefore, the power of the first can not be used, whereas the power of the second can be used for the performance of useful work. 

Although any dc power observed under equilibrium conditions is a challenge to the second law, most scientists prefer to disregard the problem connected with numerous observations of a persistent current at $R > 0$. Some scientists state that the persistent current does not threaten the second law since it is an equilibrium phenomenon and therefore no work can be extracted from the persistent current. Indeed, the free energy $F = E - ST$ has a minimum value in the equilibrium state, and no one can decrease the value below its minimum. But, the internal energy $E$ can be decreased without a decrease of the free energy if the entropy $S$ decreases at the same time. Thus, this statement can be restated as: \textit{the second law can not be broken since it can not be broken}. A philosopher noted \cite{43} that the arguments of defenders of the second law are circular as often as not.

\bigskip

\noindent
\textbf{8. Experimental evidence of the intrinsic breach of right-left symmetry}

\bigskip

\noindent
Because of the belief in the second law, some authors consider the persistent current as nondissipative \cite{36} whereas I.O. Kulik, who first predicted a persistent current in a normal metal mesoscopic loop as early as 1970 \cite{44}, wrote that this current can be observed at non-zero dissipation. Some scientists assume that the persistent current is not quite a real current and the coexistence of a finite Ohmic resistance and the equilibrium dc current is not paradoxical when one properly takes into account the influence of the measuring leads \cite{45}. But, experimental results \cite{46,47,48} prove that the persistent current behaves like a conventional current at least in a superconductor loop. 

Any circular direct current can break only clockwise - counter-clockwise symmetry in a symmetric loop. For example, there can be only a circular electric field $E = -dA/dt$ and any potential difference $V$ can not be observed on any segment because of the symmetry when the conventional circular current $I = - R_{l}^{-1}d\Phi /dt$ is induced by the Faraday voltage $-d\Phi /dt$ in a conventional normal metal symmetric loop having a resistance $R_{l}$. But, the potential difference 
$$V = (\frac{R_{ls}}{l_s} - \frac{R_l}{l})l_s I \eqno{(4)}$$
should be observed at $I \neq 0$ on a segment $l_{s}$ of asymmetric loop, wherein the resistance along the segment $R_{s}/l_{s}$ differs from that along the whole loop $R_{l}/l$. $V = 0$ in a symmetric loop in which $R_{s}/l_{s} = R_{l}/l$ according to (4) and the symmetry. The potential electric field $E = -\nabla V$ has right or left direction. Thus, the Faraday voltage breaks only clockwise - counter-clockwise symmetry in a symmetric loop and both clockwise - counter-clockwise and right - left symmetry in an asymmetric one. 

\begin{figure}
\includegraphics{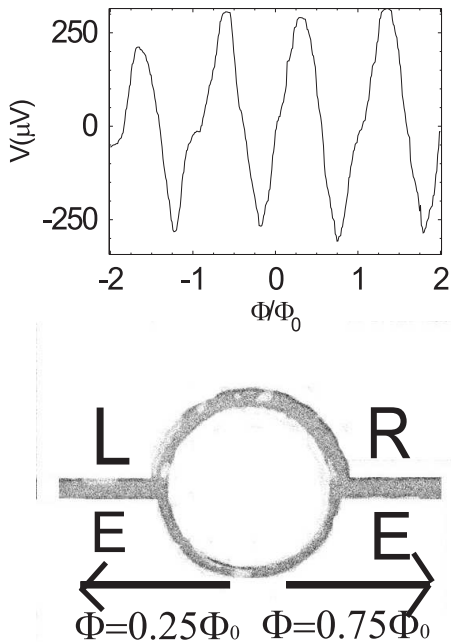}
\caption{\label{fig:epsart} Experimental evidence of the intrinsic breach of the right-left symmetry \cite{48}. The potential difference $V$ measured between points {\bf L} and {\bf R} changes sign and the electric field $E = -\nabla V$ changes direction with the scalar value $\Phi/\Phi_{0}$ at $\Phi = n\Phi_{0}$ and $\Phi = (n+0.5)\Phi_{0}$. If E has left direction at $\Phi = 0.25\Phi_{0}$ then at $\Phi = 0.75\Phi_{0}$ it has right direction.}
\end{figure}

Experiments \cite{46,47,48} show that the same is observed in the case of a persistent current in a superconductor loop with non-zero resistance. The dc potential difference, the value and sign of which are a periodical function of the magnetic field like the persistent current $V(\Phi /\Phi_{0})  \propto  I_{p}(\Phi /\Phi_{0})$, was observed in segments of an asymmetric loop \cite{46,47,48}, whereas this potential difference is not observed in segments of a symmetric loop \cite{47}. This analogy between the conventional and persistent currents can be explained by a common feature of the quantum force, maintaining the persistent current at $R > 0$ and the Faraday electromotive force. Both forces cannot be localized in any segment of the loop. The quantum force can not be localized in theory \cite{23} because of the uncertainty principle. The momentum circulation changes because of the quantization (1) when it has a certain quantum value, i.e., when the superconducting pairs are not localized in any loop segment. 

The dc voltage $V(\Phi /\Phi_{0})  \propto  I_{p}(\Phi /\Phi_{0})$ is proportional to the persistent current, and can only be observed if the loop is switched between superconducting states with different connectivity \cite{23,26,46,47,48} since a static persistent current can only exist in the closed superconducting state when the loop resistance is zero. The possibility of $V(\Phi /\Phi_{0})  \propto  I_{p}(\Phi /\Phi_{0})$ is obvious from a consideration of the switching of the $l_{s}$ segment between the normal and superconducting states \cite{26}, see Section 6. The average potential difference $V_{dc} =  <V(t)> = <R_{ls}I(t)> = <R_{ls}I_{p} \exp(-t/\tau_{RL})> \approx L_{l}I_{p}\omega_{sw}$ at a low frequency $\omega_{sw} \ll  1/\tau_{RL}$ and $V_{dc} \approx <R_{ls}>I_{p}$ at $\omega_{sw } \gg 1/\tau_{RL}$ should be observed both on the segment $l_{s}$ switched with the frequency $\omega_{sw}$ and the segment $l-l_{s}$ remaining in the superconducting state. According to the latter, a relation, like to the Josephson one \cite{49} at $\omega_{sw } \ll 1/\tau_{RL}$ 
$$V_{dc} = \frac{\pi \hbar \omega _{sw} }{e}( < n > - \frac{\Phi }{\Phi _0})\frac{l - l_s }{l} \eqno{(5)}$$
should be between the dc voltage $V_{dc}$ and the switching frequency $\omega_{sw}$, since only the electric force $2e<E> = 2eV_{dc}/(l-l_{s})$ and the quantum force $F_{q} =2\pi \hbar (<n>- \Phi /\Phi_{0})\omega_{sw}/l$ act on pair in the segment $l-l_{s}$. 

The possibility of the dc voltage on the long superconductor segment seems strange for many scientists, nevertheless, it is corroborated by experimental results \cite{48}. The measurements \cite{48} of the potential difference $V_{dc}$ between points $L$ and $R$ (see Fig.1) on a thin-walled superconducting loop, like that shown at Fig. 1, as a function of the external current $I_{ex}$ between $L$ and $R$, have shown that the current-voltage curves $V(I_{ex})$ change periodically with magnetic field at $T < T_{c}$. This periodical variation is explained by superposition of an external $I_{ex}$ and persistent $I_{p}$ currents \cite{48}. The circular persistent current increases the total current in one loop half $I_{n} = I_{ex}s_{n}/(s_{w} + s_{n}) + I_{p}$ and decreases it in the other one $I_{w} = I_{ex}s_{w}/(s_{w} + s_{n}) - I_{p}$. Here $s_{w}$ and $s_{n}$ are the sectional areas of different loop halves, see Fig. 1. The loop half should switch in the resistive state with $R_{n} > 0$ when the density of the total current $j_{n} = I_{ex}/(s_{w} + s_{n}) + I_{p}/s_{n}$ exceeds the critical value $j_{c}$. But this resistive state cannot be stable until $I_{ex}/(s_{w}+s_{n}) < j_{c}$ since the persistent current $I_{p}$ should decrease down to zero at $R_{n} > 0$. Therefore, the loop should switch between superconducting states with different connectivity in the value region of external current $(s_{w} + s_{n})(j_{c} - I_{p}/s_{n}) < I_{ex} < (s_{w} + s_{n})j_{c}$ with an intrinsic frequency $\omega_{sw}$ determined by the time of the relaxation to the equilibrium superconducting state. The other loop half with the current density $j_{w} = I_{ex}/(s_{w} + s_{n}) - I_{p}/s_{n} < j_{c}$ remains constantly in the superconducting state at $I_{ex} < (s_{w} + s_{n})j_{c}$. A dc voltage exceeding $1 \ mV$ was observed on a superconducting strip with a length $160 \ \mu m$ in a dynamic resistive state of a system of 20 loops \cite{48}, i.e. $V_{dc} > 50 \ \mu V$ on each loop. This value corresponds to the switching frequency $\omega_{sw} > 60 \ GHz$ \cite{48}. 

The asymmetry of the current-voltage curves of asymmetric loops with $s_{w} \neq s_{n}$ is observed at $\Phi \neq n\Phi_{0}$ and $\Phi \neq (n+0.5)\Phi_{0}$, since the critical values, $|I_{ex}| _{c+}=(s_{w} + s_{n})j_{c }- I_{p}/s_{n}$ and $|I_{ex}| _{c-}=(s_{w} + s_{n})j_{c }- I_{p}/s_{w}$, of the external current depends on its direction at $I_{p} \neq 0$. The value and sign of this asymmetry are periodical functions of the magnetic flux because of the $I_{p}(\Phi /\Phi_{0})$ periodical dependence. 

This dependence of the asymmetry on the scalar value $\Phi /\Phi_{0}$ and the observation of the quantum oscillation of the dc voltage $V(\Phi /\Phi_{0})$ \cite{46,47,48} are experimental evidence of the intrinsic breach of right - left symmetry. The right - left symmetry is broken in the loops both with conventional and the persistent circular currents when the external breach of symmetry (the loop asymmetry) is added to the breach of clockwise - counter-clockwise symmetry. The breach of the symmetry in the first case is external since right or left direction of the potential electric field $E = -\nabla V$ is determined by the direction of the circular Faraday electric field $E = -dA/dt$ and the external loop asymmetry. The dc voltage $V(\Phi /\Phi_{0})$ in the quantum oscillation phenomenon \cite{46-48} is observed in a constant magnetic field, without the circular Faraday electric field, and the direction of $E = -\nabla V$ changes with the scalar value $\Phi /\Phi_{0 }$ without an external vector factor: if $E$ has left direction at $\Phi  = 0.25\Phi_{0 }$ then at $\Phi  = 0.75\Phi_{0 }$ it has right direction, see Fig. 1. This intrinsic breach of right - left symmetry is a direct consequence of the intrinsic breach of the clockwise - counter-clockwise symmetry observed in the persistent current phenomenon. 

\bigskip

\noindent
\textbf{9. Actual Violation of the Second Law}

\bigskip

\noindent
In conformity with the logical consideration of Section 2, the circular persistent current breaks the clockwise - counter-clockwise symmetry without violation of equilibrium conditions and this equilibrium phenomenon can only be regarded as a potential violation of the second law. The intrinsic breach of the right - left symmetry observed in the phenomena of the dc voltage quantum oscillation $V(\Phi /\Phi_{0})$ \cite{46,47,48} clears the way for actual violation of the second law. 

\begin{figure}
\includegraphics{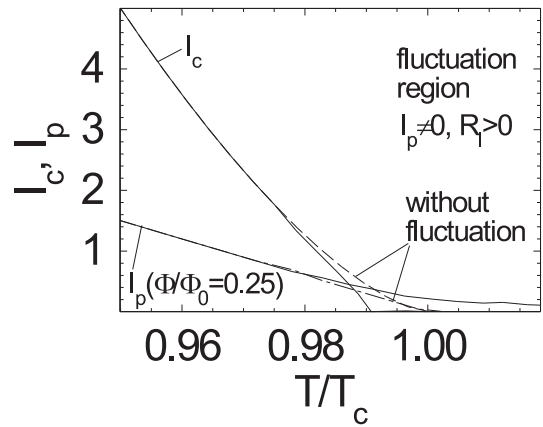}
\caption{\label{fig:epsart} According to the theory, disregarding thermal fluctuations, the persistent current and critical current diminish down to zero simultaneously at $T = T_{c}$. Because of thermal fluctuations $I_{c} = 0$, i.e. $R > 0$, and $I_{p} \neq 0$ near $T_{c}$ and, therefore, the quantum oscillations of the dc voltage $V(\Phi/\Phi_{0})$ can be observed under equilibrium conditions in the fluctuation region.}
\end{figure}

According to the quantum oscillations experiment \cite{46,47,48}, a segment of an asymmetric superconducting loop is a source of dc power $W_{dc} =V_{s}^{2}(\Phi /\Phi_{0})/R_{s}$ with a finite internal resistance $R_{s} \le R_{sn}$ at a constant magnetic flux $\Phi \neq n\Phi_{0}$ and $\Phi \neq (n+0.5)\Phi_{0}$. It is well known that, if a dc power source with a voltage $V_{s}$ and an internal resistance $R_{s}$ is loaded with a device having a resistance $R_{L}$, an amount of work or energy is extracted by the device at a power $W_{L} = V_{s}^{2}R_{L}/(R_{L}+R_{s})^{2}$. An energy $tW_{L}$ is extracted when making the measurement since both the voltmeter and the loop segment have a finite resistance $R_{L}$ and $R_{s}$. The energy $tW_{L}$ extracted in this case can be of any large magnitude since the time $t$ can be arbitrarily long. 

The energy can be extracted both in the non-equilibrium and equilibrium cases since the frequency spectrum of the dc power, $W_{dc} \neq 0$ at $\omega  =0$, differs strongly from the frequency spectrum of the equilibrium power $W_{Nyq} = k_{B}T\Delta \omega $ of other elements of the electric circuit, see Section 2. An existence of a source of the persistent power, which can be used in a device, means an actual violation of the second law in all formulations: if the device is an electric motor, then useful work can be performed, contrary to Thomson's formulation (the Carnot's principle) and the total entropy may be systematically reduced when the electric motor revolves a fly-wheel; if the device is an electric heater at a high temperature, then the heat energy can be transferred from a cold body (the dc power source) to a hot body (the heater) without an expenditure of additional energy, contrary to the Clausius formulation. 

According to the experimental results obtained in \cite{47,48} quantum oscillations of the dc voltage $V(\Phi /\Phi_{0})$ can be observed both under non-equilibrium and equilibrium conditions. Below the fluctuation region, where the loop resistance $R = 0$ under equilibrium conditions, they can be induced only by non-equilibrium noise \cite{46} or an external ac current \cite{48}. The measurements \cite{48} show that the ac current $I_{ac} = A_{I} \sin(2\pi ft)$ with any frequency $f$, from zero up to the quantum limit can induce the dc voltage $V(\Phi /\Phi_{0})$ when the current amplitude $A_{I}$ exceeds the critical value $A_{Ic} \approx (s_{w} + s_{n})j_{c }- I_{p}/s_{n}$ close to the superconducting critical current $I_{c} =  (s_{w} + s_{n})j_{c}$ of the loop. The dependence of the amplitude $A_{V} = V(\Phi /\Phi_{0} \approx 0.25)$ of the quantum oscillations $V(\Phi /\Phi_{0})$ (see Fig. 1) on the current amplitude $A_{I}$ is not monotonous since only the dynamic resistive state can make a contribution to the dc voltage $V(\Phi /\Phi_{0}) \propto I_{p}(\Phi /\Phi_{0})$ \cite{48}. The amplitude $A_{V}$ mounts a maximum value $A_{Vm}$ at $A_{I}$ slightly higher $A_{Ic}$ and decreases further with the $A_{I}$ increase \cite{48}.

Both the critical amplitude $A_{Ic}$ of the ac current and the maximum amplitude $A_{Vm}$ of the quantum oscillation $V(\Phi /\Phi_{0})$ decrease with approach to the critical temperature, $T  \to T_{c}$, since $A_{Ic} \approx I_{c}$ and $A_{Vm} \propto I_{pm}=I_{p}(\Phi /\Phi_{0} \approx 0.25)$. Without taking into account thermal fluctuations, the critical current $I_{c} \propto (1-T/T_{c})^{3 / 2}$ and the persistent current $I_{pm} \propto (1-T/T_{c})$ diminish down to zero simultaneously at $T = T_{c}$, see Fig. 2. The thermal fluctuations decrease $I_{c}$ and increase $I_{p}$, see Fig. 2. Therefore, the Little-Parks oscillations of the loop resistance $R(\Phi /\Phi_{0})$ \cite{39,40} can be observed in the fluctuation region \cite{23} where $I_{c} = 0$, i.e. $R > 0$, and $I_{p} \neq 0$ under equilibrium conditions. The resistance of the loop $R(T)$ near the superconducting transition $T  \approx  T_{c}$ is higher than zero and lower than the resistance $R_{n}$ in the normal state, i.e. at $T \gg  T_{c}$, and $I_{p} \neq 0$ at $0 < R(T) < R_{n}$ since thermal fluctuations switch segments of the loop between superconducting and normal states. Just because of such switching the quantum oscillations of the dc voltage observed on a segment of an asymmetric superconducting loop \cite{46,47,48}. Therefore, it is obvious that the dc voltage $V(\Phi /\Phi_{0})$ can be observed in the fluctuation region without any external power source. 

The quantum oscillations $V(\Phi /\Phi_{0})$ without any evident power source were first observed in 1967 \cite{46} on a double point Josephson contact and, later, \cite{47} on segments of an asymmetric aluminum loop in the superconducting fluctuation region. The authors \cite{46} assumed, and have shown, that the dc voltage $V(\Phi /\Phi_{0})$ they observed is induced by an external non-equilibrium noise. Indeed, the noise power $W_{noise}$ of a real electric circuit can exceed the Nyquist's power $W_{Nyq}$ in our noisy world. One may define a temperature of non-equilibrium noise with a power $W_{noise}$ in a frequency band $\Delta \omega $ by: $T_{noise} = W_{noise}/k_{B}\Delta \omega > T$. The measurements of the quantum oscillations of the aluminum loops, for example, are performed at $T \approx 1.2 \ K$ with help of a helium cryostat, whereas the measuring equipment has room temperature $T \approx 300 \ K$. Therefore, the noise temperature $300 > T_{noise} > 1.2$ of the loops exceeds the equilibrium temperature $T \approx 1.2 \ K$  in some band $\Delta \omega $ of the frequency spectrum. It is practically impossible to ensure equilibrium conditions in this situation. According to the results obtained in \cite{48}, a noise with any frequency from $\omega  =0$ up to the quantum limit can induce the quantum oscillations $V(\Phi /\Phi_{0})$ when its current amplitude exceeds the critical value $A_{Ic}$. Therefore, the dc voltage $V(\Phi /\Phi_{0})$ observed at $T < T_{c}$ \cite{47} is induced by non-equilibrium noise.  

But, there is not a qualitative difference between equilibrium and non-equilibrium noise relative to the actual violation of the second law since the breach of the right - left symmetry observed in the quantum $V(\Phi /\Phi_{0})$ phenomenon is intrinsic in contrast to what takes place, for example, in the case of noise rectification with help of a diode \cite{33,34}. A diode can rectify only non-equilibrium noise since it cannot break the symmetry of equilibrium motion. In contrast to this, the intrinsic breach of clockwise - counter-clockwise symmetry of equilibrium motion is observed in the persistent current phenomenon. Any noise, both equilibrium and non-equilibrium, switches only the asymmetric loop in the resistive state. Therefore, this result \cite{47} is experimental evidence of an actual violation of the second law in spite of the fact that the dc power observed in this work can be induced by non-equilibrium noise. The intrinsic breach of the right - left symmetry observed in the quantum oscillation phenomena \cite{46,47,48} is experimental evidence of this actual violation. 

The authors \cite{46} assumed that the dc voltage can be induced only by non-equilibrium noise because of a belief in the second law. But also they could not be well-informed about superconducting fluctuations since the basic works by Aslamazov, Larkin \cite{50} and Maki, Thompson \cite{51} concerning this problem were published after 1967. Recently, Jorge Berger has shown \cite{52} that both non-equilibrium and equilibrium noise can induce quantum oscillations in a superconducting loop with two asymmetric Josephson junctions. 

\bigskip

\noindent
\textbf{10. Violation of the Second Law is the Most Obvious Consequence of Quantum Mechanics at the Macroscopic Level}

\bigskip

\noindent
It is difficult to believe in violation of the second law even with experimental evidence. But, it can be understood. Richard Feynman stated that he \textit{can safely say that today understands quantum physics}. But violation of the second law of thermodynamics is a most ordinary and obvious consequence of quantum physics. It is drawn from our experience of a discrete spectrum of permitted states of some quantum systems and perpetual equilibrium motion. Each physicist knows that each element of an electric circuit is power source because of thermal fluctuations and this power can be used under equilibrium conditions when the intrinsic breach of right-left symmetry is observed. Any physicist easily understands this. 

The situation becomes more difficult to understand when one considers not thermal fluctuations, but rather, quantum fluctuations. Experiments show that near the superconducting transition, i.e. at $T  \approx  T_{c}$, the average value of velocity squared $<v_{s}^{2}>$ of superconducting pairs in a loop with weak screening $LI_{p} < \Phi_{0}$ has a maximum value \cite{37,39,40}; whereas, the average velocity and the persistent current equal zero, $<v_{s}> = 0$, $I_{p}(\Phi /\Phi _{0}) \propto V(\Phi /\Phi_{0}) = 0$, \cite{47,48} at $\Phi = (n+0.5)\Phi_{0}$ although the state with $v_{s} = 0$ is forbidden. The $<v_{s}^{2}>$ value is maximum and $<v_{s}> = 0$ at $\Phi =(n+0.5)\Phi_{0}$ and $T \approx T_{c}$ since thermal fluctuations switch the superconducting loop between the permitted states $n - \Phi /\Phi_{0} = 0.5$ and $n - \Phi /\Phi_{0} = -0.5$ with the same minimum energy $ \propto v_{s}^{2} \propto (n - \Phi /\Phi_{0} )^{2}$: $<v_{s}^{2}> \propto  0.5^{2} + (-0.5)^{2} \neq  0$ and $<v_{s}>  \propto  0.5 + (-0.5) = 0$. 

According to \cite{53}, the magnetic dependence of the persistent current of the superconducting loop at low temperature $T \ll  T_{c}$ can be like the one at $T \approx T_{c}$ because of quantum fluctuations. But, it is not clear how the persistent current can be non-zero $j_{p} \neq 0$ at $\Phi  \neq (n+0.5)\Phi_{0}$ and zero $\Phi  = (n+0.5)\Phi_{0}$ in the case of quantum fluctuation. The velocity cannot change in time in the case of quantum fluctuations, since this change should induce the Faraday's voltage $d\Phi /dt = L dI_{p}/dt = Lsdj_{p}/dt = Ls2en_{s} \ dv/dt$ which should be accompanied by an exchange of energy with the environment. 

It is assumed that quantum superposition of states can exist in a superconducting loop at $\Phi  = (n+0.5)\Phi_{0}$ like the one observed in microscopic systems. If this assumption is correct, a non-zero value of the persistent current corresponding to the permitted states $n - \Phi /\Phi_{0} = \pm 0.5$ should be observed at $\Phi  = (n+0.5)\Phi_{0}$ and the value $j_{p} = 0$ cannot be observed. But, could the quantum superposition be observed at the macroscopic level where the impossibility of noninvasive measurability raises doubts? This question is raised and discussed in the papers \cite{54,55,56,57}.

\bigskip

\noindent
\textbf{11. Conclusions }

\bigskip

\noindent
Most scientists are fully confident in the absolute status of the second law, a view held for more than one and one-half centuries, notwithstanding that this belief presents a difficulty for the scientific explanation of the world. If total chaos can only increase, then why is any order observed? No scientist can explain it conclusively, although many words are said about this. The possibility of violation of the second law because of the intrinsic breach of symmetry in quantum systems removes this difficulty, but I am not sure that most scientists can tolerate this disproof of the demand that it is irreversible that total entropy must increase, based on the centuries-old belief. 

One of the most obvious difficulties for the belief in irreversible entropy increase is connected with living systems. The total entropy increases when we use petrol in our cars. This process is opposite to the decrease of the entropy in very old living systems. Because of the belief in the second law, most scientists are fully confident that this entropy decrease can only be local. But nobody can explain how the Carnot principle can work on the molecular level where the appeal to \textit{an immense number of molecules} is not valid. A real discovery of a mechanism of violation of the second law in living systems could resolve this obvious problem. One of the possible mechanisms was found by Vladislav Capek in the interaction between ions and biomolecules \cite{4}. He stated \cite{2} that those breakdowns of the second law might be occurring in living systems. The result \cite{4} as well as the other works \cite{8,9,10,11,12} by Professor Capek will be appreciated for its true value in the future. 

The possibility of an actual violation of the second law, i.e. of the Carnot principle, means that cars can move without any fuel. The persistent power cannot exceed the total power of thermal fluctuations $W_{fl} \approx  (k_{B}T)^{2}/\hbar $ \cite{22,23,26} which is weak: $W_{fl} \approx 10^{ - 12} \ Wt$ at $T = 1 \ K$ and $W_{fl} \approx 10^{ - 8} \ Wt$ at $T = 100 \ K$. Therefore, it is important that the dc power can be summed in contrast to the Nyquist's noise. It is well known that the Nyquist's power of one resistor $R$ equal the one of $N$ resistors $W_{Nyq} = k_{B}T\Delta \omega $ whereas the voltage of $N$ dc power sources connected in series equals $NV$ and the maximum power in a load $W_{L} = N^{2}V_{s}^{2}R_{L}/(R_{L}+NR_{s})^{2} = NV_{s}^{2}/4R_{s}$ at $R_{L} = NR_{s}$ is in $N$ time higher for the $N$ dc power sources than for single one, $W_{L} = V_{s}^{2}/4R_{s}$ at $R_{L}= R_{s}$. This difference is the consequence of the difference between chaotic power and ordered power. Thus, persistent power can be made very large. The only limitation may be technological problems. The first results \cite{48} have shown that it is easy enough to make a system of asymmetric superconductor loops connected in series in which the dc voltages are summed. From the very first we have obtained on a system of 20 loops the quantum oscillations with amplitude approximately 20 times higher than that of a single loop under the same conditions \cite{48}. 

\bigskip

\noindent
\textbf{Acknowledgements }

\bigskip

I thank Jorge Berger for sending me the paper \cite{46}. This work was financially supported by the Russian Foundation of Basic Research (Grant 04-02-17068), the Presidium of Russian Academy of Sciences in the Program "Low-Dimensional Quantum Structures" and ITCS department of Russian Academy of Sciences in the Program "Technology Basis of New Computing Methods".

\end{document}